\documentclass[twocolumn]{aastex62}

\newcommand{\beq}{\begin{equation}}
\newcommand{\eeq}{\end{equation}}

\usepackage{ulem}

\def \degmark{^\circ}

\def \phcmsec{\hbox{ph cm$^{-2}$ s$^{-1}$}}
\def \ctscmsecsr{\hbox{cts cm$^{-2}$ s$^{-1}$ sr$^{-1}$}}

\def \cmsec{\hbox{cm$^{2}$ s}}

\def \gray {$\gamma$-ray }
\def \grays {$\gamma$-rays }

\graphicspath{{./}{figures/}}

\accepted{April 19, 2021}
\submitjournal{ApJ}

\shorttitle{CNN for AGILE-GRID GRBs Detection}
\shortauthors{N. Parmiggiani et al.}
\begin{document}

\title{A Deep Learning Method for AGILE-GRID GRB Detection}

\correspondingauthor{N. Parmiggiani}
\email{nicolo.parmiggiani@inaf.it}

\author[0000-0002-4535-5329]{N. Parmiggiani}
\affiliation{INAF OAS Bologna, Via P. Gobetti 93/3, 40129 Bologna, Italy}
\affiliation{Universit\'a degli Studi di Modena e Reggio Emilia, DIEF - Via Pietro Vivarelli 10, 41125 Modena, Italy}

\author[0000-0001-6347-0649]{A. Bulgarelli}
\affiliation{INAF OAS Bologna, Via P. Gobetti 93/3, 40129 Bologna, Italy}

\author[0000-0002-6082-5384]{V. Fioretti}
\affiliation{INAF OAS Bologna, Via P. Gobetti 93/3, 40129 Bologna, Italy}

\author[0000-0002-9894-7491]{A. Di Piano}
\affiliation{INAF OAS Bologna, Via P. Gobetti 93/3, 40129 Bologna, Italy}

\author[0000-0002-4315-1699]{A. Giuliani}
\affiliation{INAF-IASF Milano, Via Alfonso Corti 12, I-20133 Milano, Italy.}

\author[0000-0003-2501-2270]{F. Longo}
\affiliation{Dipartimento di Fisica, University of Trieste, via Valerio 2, 34127 Trieste, Italy}

\affiliation{INFN, sezione di Trieste, via Valerio 2, 34127 Trieste, Italy}
\affiliation{Institute for Fundamental Physics of the Universe Via Beirut 2, Trieste, Italy}

\author[0000-0003-3455-5082]{F. Verrecchia}
\affiliation{ASI Space Science Data Center (SSDC), Via del Politecnico snc, 00133 Roma, Italy.}

\affiliation{INAF-Osservatorio Astronomico di Roma, Via di Frascati 33, I-00078 Monte Porzio Catone, Italy.}

\author[0000-0003-2893-1459]{M. Tavani}

 \affiliation{INAF-IAPS Roma, via del Fosso del Cavaliere 100, I-00133 Roma, Italy.}
 \affiliation{Dipartimento di Fisica, Universit\'a Tor Vergata, via della Ricerca Scientifica 1,I-00133 Roma, Italy.}
\affiliation{INFN Roma Tor Vergata, via della Ricerca Scientifica 1, I-00133 Roma, Italy}
\affiliation{Consorzio Interuniversitario Fisica Spaziale (CIFS), villa Gualino - v.le Settimio Severo 63, I-10133 Torino, Italy.}

\author[0000-0001-6616-1753]{D. Beneventano}
\affiliation{Universit\'a degli Studi di Modena e Reggio Emilia, DIEF - Via Pietro Vivarelli 10, 41125 Modena, Italy}

\author[0000-0002-1348-250X]{A. Macaluso}
\affiliation{University of Bologna  - Department of Computer Science and Engineering (DISI)
Viale del Risorgimento 2, 40136 Bologna, Italy}



\begin{abstract}
The follow-up of external science alerts received from Gamma-Ray Bursts (GRB) and Gravitational Waves (GW) detectors is one of the AGILE Team's current major activities. The AGILE team developed an automated real-time analysis pipeline to analyse AGILE Gamma-Ray Imaging Detector (GRID) data to detect possible counterparts in the energy range 0.1-10 GeV. This work presents a new approach for detecting GRBs using a Convolutional Neural Network (CNN) to classify the AGILE-GRID intensity maps improving the GRBs detection capability over the Li\&Ma method, currently used by the AGILE team. The CNN is trained with large simulated datasets of intensity maps. The AGILE complex observing pattern due to the so-called 'spinning mode' is studied to prepare datasets to test and evaluate the CNN. A GRB emission model is defined from the Second Fermi-LAT GRB catalogue and convoluted with the AGILE observing pattern. Different \textit{p-value} distributions are calculated evaluating with the CNN millions of background-only maps simulated varying the background level. The CNN is then used on real data to analyse the AGILE-GRID data archive, searching for GRB detections using the trigger time and position taken from the Swift-BAT, Fermi-GBM, and Fermi-LAT GRB catalogues. From these catalogues, the CNN detects 21 GRBs with a significance $\geq 3 \sigma$, while the Li\&Ma method detects only two GRBs. The results shown in this work demonstrate that the CNN is more effective in detecting GRBs than the Li\&Ma method in this context and can be implemented into the AGILE-GRID real-time analysis pipeline.
\end{abstract}

\keywords{Gamma-rays, Gamma-ray bursts, Neural networks, Convolutional neural network}


\section{Introduction \label{sec:intro}}

AGILE (Astrorivelatore Gamma ad Immagini LEggero Light Imager for Gamma-Ray Astrophysics) is a scientific mission of the Italian Space Agency (ASI) that was launched on 23rd Apr 2007 \citep{Tavani,Tavani1}. The AGILE payload detector consists of the Silicon Tracker (ST) \citep{2001AIPC..587..754B, 2003NIMPA.501..280P, bulgarelli10, cattaneo11} the SuperAGILE X-ray detector \citep{2007NIMPA.581..728F}, the CsI(Tl) Mini-Calorimeter (MCAL) \citep{labanti09}, and an AntiCoincidence System (ACS) \citep{2006NIMPA.556..228P}. The combination of ST, MCAL, and ACS forms the Gamma-Ray Imaging Detector (GRID). The AGILE-GRID is used for observations in the 30 MeV-50 GeV energy range. The Precise Positioning System and the two Star Sensors provide accurate timing, positional, and attitude information. The ST is the core of the AGILE-GRID, and it relies on the process of photon conversion into electron-positron pairs. It consists of 12 trays, the first 10 of which include a tungsten converter followed by a pair of silicon microstrip detectors with strips orthogonal to each other, the last two consisting only of silicon detectors. The \grays are converted in the tungsten (silicon) layers, and a readout electronics acquire and process the data.

The AGILE team developed an automated pipeline to react to external science alerts received from the Gamma-Ray Coordinates Network (GCN) \citep{bulgarelli2019b}. A science alert is a communication from/to the astrophysical community that a transient phenomenon occurs in the sky. This automated pipeline can react in a fast way and detect a possible GRB counterpart in AGILE-GRID short term ($<1000$ sec) observations. A detection occurs when the pipeline finds a signal with a statistical significance above a defined threshold. The GRB positions and the trigger times are known in advance because the pipeline reacts to external science alerts. The analysis is performed using aperture photometry, evaluating the counts detected inside a time window containing the target ($T_{on}$) and a time window containing only background ($T_{off}$). The AGILE-GRID instrument has a Point Spread Function (PSF) $<10 \degmark$ at energy $>50$ MeV \citep{Sabatini2015}. The counts are selected from the AGILE-GRID photon list in a radius of $10 \degmark$ from the centre of the error localization region reported by the external science alert in order to contain the PSF of the source. The background is evaluated before the trigger time because the true duration of the GRB in the GRID energy range is unknown. 

In the current method used by the AGILE-GRID automated pipeline, the significance of a GRB detection is calculated with the Li\&Ma formula \citep{li83} using the counts extracted in the previous steps. The Li\&Ma is a likelihood ratio method applied to aperture photometry. It is largely used in \gray astronomy and by the AGILE Team as a standard analysis for GRB detection. This method, cited from now on as Li\&Ma, has two main limits.

The Li\&Ma does not use the shape of the PSF during the analysis, but just the event number inside a region defined large enough to include the PSF. Furthermore, the Li\&Ma method requires the counts in both $T_{on}$ and $T_{off}$ time windows to be not too few, with a threshold of ten usually applied \citep{li83}. The AGILE team, following the \cite{li83}, defined the threshold of ten counts for the real-time analysis pipeline, and for this reason, detections with a lower count rate are discarded.

This work proposes a new detection method to overcome these limits and, in general, improve the AGILE-GRID automated pipeline's capability to detect GRBs during the follow-up of science alerts received from other observatories through the GCN network.

This new method uses a class of Deep Learning (DL) methods called Convolutional Neural Network (CNN) described in detail in Sect. \ref{sec:cnn}. 

The DL methods \citep{LeCun2015} are part of the Machine Learning (ML) methods. The ML methods use automated training algorithms (without human intervention) to learn how to predict the correct output concerning several problems (classification, regression, etc.) without being directly programmed to do this. The training is performed using a training dataset that is a subset of the whole population of possible inputs that the model will obtain to predict the output. The ML techniques can not be used directly on the raw data but require a first step of feature engineering (feature extraction) from the raw data. This operation is time-consuming and must be performed by field experts with a complete understanding of the data. Once extracted, the features are used as input for the ML model. The DL methods, on the contrary, do not require this feature engineering performed by experts because they can extract features directly from the raw data. The DL architectures, called Deep Neural Network (DNN), are composed of several layers that are able to extract features at different levels of abstraction during the training phase. The number of layers can vary with the problem complexity and the available computing power starting from less than ten layers \citep{Krizhevsky2012} to more than one hundred \citep{He2015} and further. The DNNs have become even more popular in recent years thanks to three main factors: (i) the improvement of computational hardware (e.g. Graphical Processing Unit - GPU) required to train DNN with millions or billions of parameters, (ii) the availability of huge amounts of data suitable for the training of large DNN models and, (iii) the development of frameworks that can be used to implement these DNN models with standard technologies (e.g. Python).

The CNN developed in this work is used to classify AGILE-GRID intensity maps and detect the presence of GRBs in the field. Intensity maps are counts maps divided by the exposure, and therefore report the measurement in $\phcmsec sr^{-1}$ for each pixel. The CNN uses the intensity maps as input and does not require information about the exposure.
 
The CNN requires a training phase with large simulated datasets of intensity maps representing the average background level and the GRB flux distribution expected in the AGILE-GRID energy range 0.1-10 GeV.  The study of the observing pattern is described in Sect. \ref{sec:obs_model}. Sect. \ref{sec:grb} describes the GRB model used to simulate GRBs for the CNN training. After performing the CNN training, the \textit{p-value} distribution from only background maps is computed in different observational conditions (Sect. \ref{sec:far}). The CNN is then applied to real data using the GRBs' position and trigger time of the Swift-BAT\footnote{Swift-BAT Gamma-Ray Bursts Online catalogue https://swift.gsfc.nasa.gov/archive/grb\_table/}, Fermi-LAT \citep{ajello2019} and Fermi-GBM \footnote{Fermi-GBM Gamma-Ray Bursts catalogue https://heasarc.gsfc.nasa.gov/W3Browse/fermi/fermigbrst.html} catalogues. Sect. \ref{sec:searchgrb} describes this analysis, and the results show a considerable improvement in the detection capability compared to the Li\&Ma.

The main reasons why the CNN method improves the AGILE-GRID GRBs detection capabilities are:
 \begin{enumerate}
    \item The CNN can be trained on the data of a specific instrument, learning from huge datasets of simulated data, while the Li\&Ma is a generic method. In fact, the PSF of the AGILE-GRID instrument is used during the CNN training phase to define the size of the kernels used during the convolution process.
    \item  The CNN is trained with datasets simulated using the background level calculated during real AGILE-GRID observation. In addition, the fluxes of the simulated GRBs are extracted from the Fermi-LAT GRB catalogue \citep{ajello2019} and scaled to the AGILE energy range. All this knowledge is learned by the CNN, while the Li\&Ma is applied as-is.
    \item  The CNN does not require a minimum number of events to be applicable. On the contrary, the Li\&Ma requires at least ten events in the $T_{on}$ and $T_{off}$ time windows. 
 \end{enumerate}
 
This is the first attempt to use a CNN to classify the AGILE-GRID \gray sky maps. The results obtained in this work (Sect. \ref{sec:searchgrb}) encourage further researches in this direction. Future works are planned to use the CNN to classify AGILE-GRID \gray sky maps containing more than one source and perform a regression analysis to determine the GRB position and the flux. These kinds of analyses can not be performed with the Li\&Ma. The method described in this work can also be used to train a CNN network to classify sky maps produced by the next generation of X-ray and \gray observatories such as the Cherenkov Telescope Array \citep{Actis2011,Acharya2019} or the e-ASTROGAM \citep{deangelis2021} and THESEUS \citep{amati2021} spacecraft. These observatories will produce more complex sky maps collecting a larger number of events and background information. The CNN can be trained on the observing condition (e.g. the background level) of the specific instrument and can learn detection patterns following the instrument's PSF. These are additional reasons to promote the research in this field.

The CNN is used in astrophysics to analyse data in several contexts. In particular, the CNN can be used for images classification problems. As described in \cite{Hezaveh_2017}, the CNN is used to perform fast and automated gravitational lenses analyses. With the next generation of ground and space observatories such as the Vera C. Ruby Observatory (formerly Large Synoptic Survey Telescope - LSST ) \citep{Thomas_2020}, tens of thousands of new lenses are expected to be discovered. The CNN can improve these analyses' performances and reduce the time required to obtain results in contrast with traditional analyses methods based on Maximum Likelihood Estimator (MLE). The CNN technologies can be used to analyse the Big Data generated by the next generation of observatories, exploiting the GPUs computing power and parallel processing. Several works use the CNN in the \gray data analysis. In \cite{Caron2018}, the CNN approach is used to analyse Fermi-LAT \gray maps of the Galactic Center. In \cite{Drozdova2020}, the CNN method is used to extract point sources on Fermi-LAT simulated images.

\section{Assumptions of this work \label{sec:work_assumptions}}

The analyses presented in this paper use parameters inherited from the AGILE real-time analysis pipeline developed for the follow-up of external science alert. The new method based on DL techniques is compared with the standard method used in this pipeline. The parameters that are not inherited by the AGILE pipeline are defined here to test the CNN model with the common conditions that can be found during the AGILE-GRID observations. This work does not treat rare and complex situations that will be analysed in future works. Not all parameters are fixed. They may be calculated during the analyses (e.g. the time window used to evaluate the background level). The main assumption made for this work are:
\begin{enumerate}
    \item The \gray sky maps used to train and evaluate the CNN have a size of $100\times100$ pixels and a bin size of $0.5\degmark$. This map size is defined to be larger than the AGILE-GRID PSF and to include background regions. The bin size is a standard parameter used in the AGILE-GRID data analysis.
    \item The \gray sky are simulated with a time window of 200 seconds. This value is selected after the analysis of the AGILE observing pattern described in Sect. \ref{sec:param_id}. 
    \item The maps are simulated using a representative value for the AGILE-GRID exposure in maps with 200 sec time windows, calculated excluding exposure levels under a threshold defined to avoid limit conditions that are not the goal for this work. 
    \item The energy range considered in this work is 0.1-10 GeV. This energy range is the standard one used by the AGILE Team to perform analysis on AGILE-GRID data and is supported by the AGILE Science Tools' simulation software.
    \item This work is focused on GRBs in the extra-Galactic region ($|b|>10\degmark$, where b is the Galactic latitude) to evaluate the new method excluding regions with several background sources and to avoid the diffuse Galactic \gray background (Sect. \ref{sec:bkg_estimation}).
    \item The background levels of the maps classified with the CNN are calculated with an MLE analysis. The AGILE Team defined that the time windows to calculate the background with an MLE analysis must contain a minimum of ten counts. The time windows are found automatically to have at least ten counts. For a minimum of 10 counts, the background time window last from 6 to 32 hours. During extra-Galactic observations, the AGILE-GRID background is isotropic and mainly dominated by charged particles populating the low earth orbit radiation environment (see Sect. \ref{subsec:agile_bkg} for a full description). This background flux is fairly quiescent, and no significant variations are expected for time scales from several hours to few days. In fact, when studying the general background fluctuation trends within a year, an averaged variability of only 30\% (1 sigma) is found.
    \item As described in \cite{bulgarelli12a}, the AGILE Team uses analysis regions with a radius of $10 \degmark$ for the AGILE-GRID data centred on the source position to include the PSF of the instrument. 
    \item The external science alerts considered for this work have a maximum error region of $1 \degmark$. This scenario covers more than 90\% of GRBs presented in the Second Fermi-LAT GRBs catalogue and 100\% of the GRBs reported in the Swift-BAT GRBs catalogue. The science alerts with greater error regions are excluded because this work does not have the goal to find the source's position in a blind search. 
    \item The CNN uses intensity maps (counts maps divided by the exposure maps) as input. This solution makes the CNN exposure-independent.
\end{enumerate}

\section{Modelling the Observations \label{sec:obs_model}} 

The AGILE orbit (quasi-equatorial with an inclination angle of $2.5\degmark$ and an average altitude of 500 km, 96 min period) is optimal for low-background \gray observations. From July 2007 to October 2009, AGILE observed the \gray sky in 'pointing mode', characterised by quasi-fixed pointing with a slow drift ($\sim 1\degmark$/day) of the instrument boresight direction following solar panel constraints. 

Due to a change in the satellite pointing control system, since November 2009, the AGILE \gray observations have been obtained with the instrument operating in 'spinning mode' (i.e., the satellite axis sweeps a $360\degmark$ circle in the sky approximately every 7 min). The axis of this circle points toward the Sun, so the whole sky is exposed every six months.

This new mission configuration provides a unique capability to the AGILE satellite to discover transients. The actual spinning configuration of the satellite, together with a large field of view and a sensitivity of typically, $F = (1-2) \times 10^{-8} \rm  \, erg \, cm^{-2} \,  s^{-1}$ for 100 sec time integration, provides a coverage of $80\%$ of the sky, with each sky position covered 200 times per day with 100 sec of integration time.

\subsection{Parameters identification} \label{sec:param_id}

\begin{figure}[!htb]
\plotone{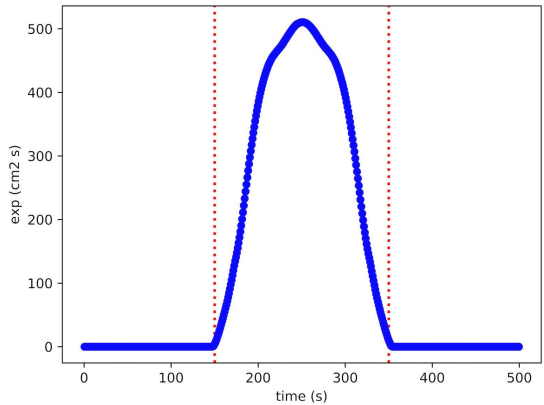}
\caption{Typical pattern of the AGILE-GRID exposure for a fixed accessible sky region given in values of [$\rm cm^2 \, s$] as a function of time during the AGILE spinning mode. The red dotted lines represent the time window of 200 sec used for the maps simulation.}
\label{fig:exp3}
\end{figure}

The complex observing pattern of AGILE in  'spinning mode'  is studied to identify the range of conditions during the observations (average exposure level and background level). These conditions are used to perform the Monte Carlo simulations of training, validation, and test datasets. Because this work is focused on extra-galactic sky regions, a sky region centred at Galactic coordinates $(l,b)=(45,30)$ is selected to study the exposure and background level. The AGILE-GRID exposure in the centre of this sky region is calculated during one spinning revolution. Fig. \ref{fig:exp3} shows the typical exposure pattern for fixed accessible sky regions: the exposure values are calculated in a time window of 500 sec divided in 1 sec bins and in a radius of $10\degmark$. The exposure is highly variable, with a well-defined peak due to the AGILE rotation. The study of the exposure pattern during the AGILE spacecraft spinning is used to determine the time window size used during the maps simulation. A time window of 200 seconds (shown in Fig. \ref{fig:exp3} with red dotted lines) is selected because it contains an entire spin of AGILE exposure. 

\begin{figure}[!htb]
\plotone{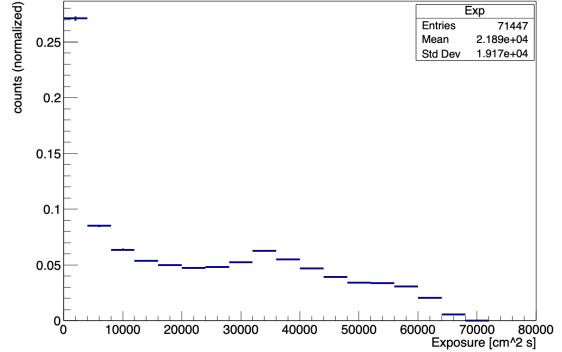}
\caption{Histogram of exposure values [\cmsec] calculated with 200 sec integrations during one year of AGILE-GRID data.}
\label{fig:exp_histo}
\end{figure}

The exposure analysis is repeated for a year of observation (1st Jan 2018, 1st Jan 2019), with the integration time of 200 sec to obtain the exposure distribution.  Excluding intervals with no exposure, the mean value obtained is $\sim 20 \times 10^3 \: \cmsec$ (Fig.\ref{fig:exp_histo}). Thus, all exposure levels lower than $20 \times 10^3 \: \cmsec$ are excluded for this work. Then a new distribution is obtained with a mean value of $\sim 40 \times 10^3 \: \cmsec$. The obtained value is used to simulate the datasets. This procedure aims to focus the training and the evaluation of the CNN on exposure values excluding limit conditions that are not the goal of this work.

\subsection{AGILE-GRID background estimation \label{subsec:agile_bkg}} 

\label{sec:bkg_estimation}
\begin{figure}[!htb]
\plotone{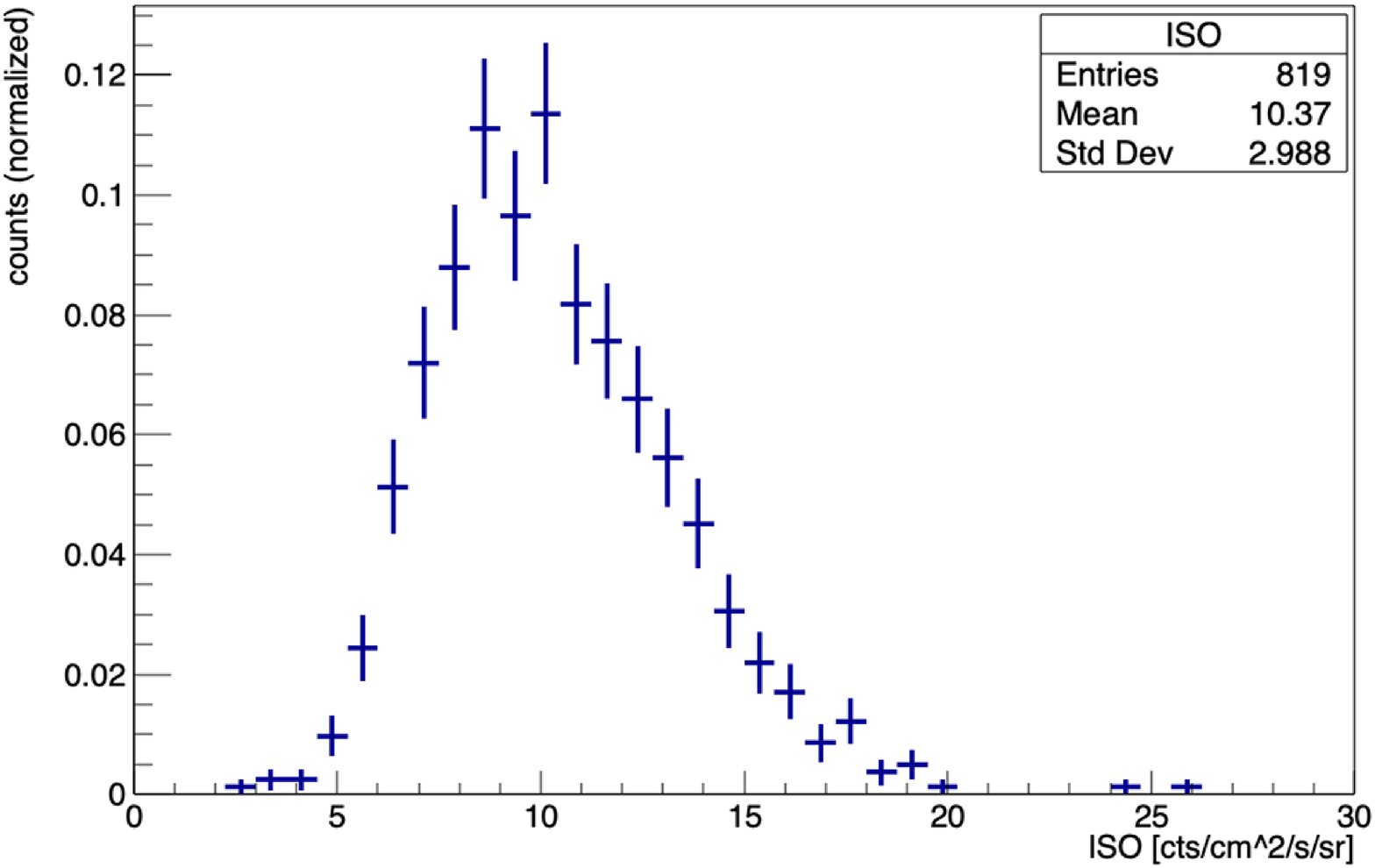}
\caption{Histogram of $g_{iso}$ values, expressed in $10^{-5} \ctscmsecsr$, for 6 hours  integrations during one year of AGILE-GRID data.}
\label{fig:iso_histo}
\end{figure}

Two background components are taken into account. The diffuse \gray background ($g_{gal}$) is assumed to be produced by the interaction of Cosmic Rays (CR) with the Galactic interstellar medium, the Cosmic Microwave Background (CMB), and the Interstellar Radiation Field (ISRF). The (quasi) isotropic background ($g_{iso}$) includes both a contribution from the cosmic extra-Galactic diffuse emission as well as a component of noise due to residual CR induced background at the detector level. In the extra-Galactic regions, the isotropic background dominates the AGILE-GRID data. For this reason, the $g_{gal}$ value is considered equal to zero. More details about the AGILE-GRID background model can be found in \cite{bulgarelli2019}. One year of data (1st Jan 2018, 1st Jan 2019) is analysed using time windows of 6 hours to obtain the distribution of $g_{iso}$ values in an extra-galactic position. The time window size is defined to have a mean of ten counts in a radius of $10 \degmark$. This counts value is required to perform the statistical analysis of the background level using the MLE method. Fig. \ref{fig:iso_histo} shows the $g_{iso}$ distribution excluding time windows with less than ten counts in a radius of $10 \degmark$. The mean of the distribution is $ 10.4 \times 10^{-5} \ctscmsecsr$ and the standard deviation is $3.0  \times 10^{-5} \ctscmsecsr$. This distribution is used to simulate the datasets to train the CNN, more detail in Sect. \ref{subsec:train_dataset_creation}.

\section{GRB Model \label{sec:grb}} 

AGILE-GRID detected so far only eleven GRBs, namely GRB080514B \citep{080514}, GRB090401B \citep{GCN9069, GCN9075}, GRB090510 \citep{090510}, GRB100724B \citep{100724}, GRB130327B \citep{GCN14344}, GRB130427A \citep{GCN14515}, GRB131108A \citep{GCN15479}, GRB170115B \citep{GCN20474}, GRB180914B \citep{GCN23231}, GRB190501A \citep{GCN24361}, and GRB1905030A \citep{GCN24683}. 

These events exhibit some of the main properties of the larger GRB population detected by the Fermi-LAT experiment and discussed, for example, in its First GRB catalogue \citep{firstGRB} and confirmed more recently in its Second GRB catalogue \citep{ajello2019}. 
From here on, the work will refer to the Second Fermi-LAT GRBs catalogue.
In particular, the GRBs' main characteristics at energies greater than 100 MeV, as detected by LAT and used in this study, are the spectral model and its temporal decay. The first one shows a clear flattening of the spectrum to a value around -2 at late times, independent of other GRB properties, and a typical larger duration concerning lower energies, extending up to 1000 sec in the First catalogue and up to 10000 sec in the second one. The temporal power-law decay index is clustered around -1. The GRB detected by the LAT, in its First catalogue, were among the brightest detected by the Fermi-GBM, with a fluence generally greater than a few $10^{-6} \rm \, erg \, cm^{-2}$ (see discussion in the First LAT GRB catalogue). In the Second LAT GRB catalogue, the fluence limit decreased up to around $10^{-6} \rm \, erg \, cm^{-2}$ for long GRBs.

\subsection{Flux estimation from LAT catalogue} \label{sec:flux_estimation}

\begin{figure}[!htb]
    \plotone{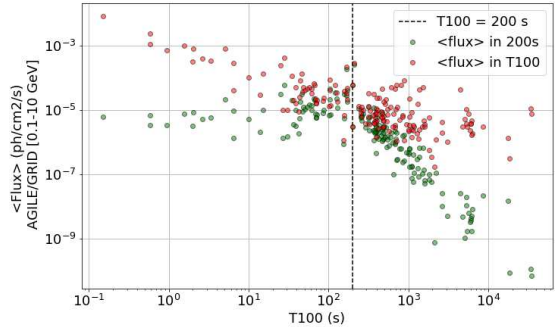}
    \caption{Average photon flux of the second Fermi-LAT GRB catalogue population as a function of the Fermi-LAT duration. Red data represent the catalogued flux values within the AGILE-GRID energy range (0.1-10 GeV). In green is the evaluation of the average flux of each event for 200 sec emission. 
    }
    \label{fig:lat-fluence}
\end{figure}

Under the assumption that the simple power-law model from \cite{ajello2019} is a good enough estimate of the spectral shape of the Fermi-LAT detected sources, the $F_{ph}^{LAT} \, (\phcmsec)$ within 0.1-100 GeV energy range observed by Fermi-LAT is scaled to a $F_{ph}^{GRID} \, (\phcmsec)$ value within the AGILE-GRID energy range (0.1-10 GeV). 
The photon fluence was then computed as $f_{ph} = F_{ph} \times T_{100}$\footnote{The duration of the burst ($T_{100}$) is defined as the time between the first and last photon detection in the 0.1-10 GeV energy range to be associated with the GRB with probability $p > 0.9$ \citep{ajello2019}.}. The fraction of the integrated photon flux emitted by the source in 200 sec exposure time is required to simulate the datasets. For events with $T_{100}$ greater than 200 sec, the photon flux emitted within the said exposure time is calculated, assuming the catalogued simple power-law evolution model to weight the loss of later emission. The average photon flux of events with $T_{100}$ less than 200 sec is instead mediated over the exposure time to preserve the total photon fluence (Fig. \ref{fig:lat-fluence}).

\begin{figure}[!htb]
    \plotone{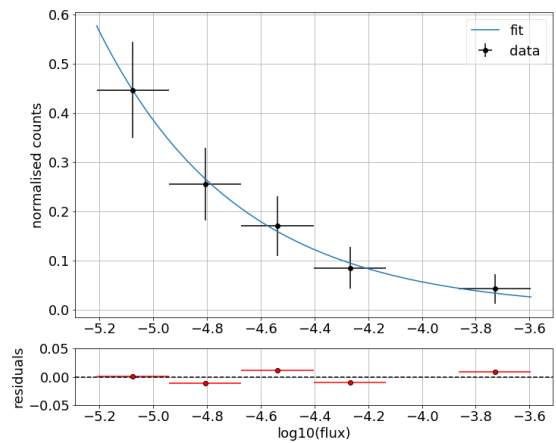}
    \caption{Gamma-ray bursts population of the second Fermi-LAT catalogue, with photon flux above $6.6\cdot10^{-6} \, \phcmsec$. The distribution fit (blue continuous line) is achieved with an exponential law. Residuals are shown in the bottom panel.}
    \label{fig:lat-fit}
\end{figure}

The simulation process is based on the use of a function fitted from GRBs data having $F > 6.6 \times 10^{-6} \, \phcmsec$ to extract random flux values. This exponential function is obtained through Levenberg-Marquardt method: $y = a \cdot e^{-x/b}$ where $x = log_{10}(F_{ph})$, $a=2.7 \times 10^{-5}$ and $b=0.523$. In Fig. \ref{fig:lat-fit} the distribution and fitting function are displayed in the top panel, with residuals shown in the bottom panel.

\section{Convolutional Neural Network \label{sec:cnn}}  

A deep learning (DL) approachm described in Sect. \ref{sec:intro}, is used to build a GRBs detection algorithm. The DL architecture used in this work is the CNN, a class of Deep Neural Network specifically developed to analyse and classify images \citep{Krizhevsky2012,Goodfellow2016}. The CNN is trained to classify the intensity maps in the AGILE-GRID energy range to detect GRBs. This CNN has a multiple layer architecture where each layer can identify specific features inside the image. The supervised learning technique is used to train the CNN. This technique requires the training of the CNN with a labelled dataset. These kinds of datasets contain the results of the classification for each element.

As described in Sect. \ref{sec:intro}, The CNN are used in several astrophysical contexts, exploiting the features of this technology for data analysis and object classification.

\subsection{Datasets simulation}
\label{subsec:train_dataset_creation}

Three Monte Carlo simulations for the training, validation, and test dataset, each with 40 000 AGILE-GRID intensity sky maps, are performed. 

\begin{figure}[!htb]
\plotone{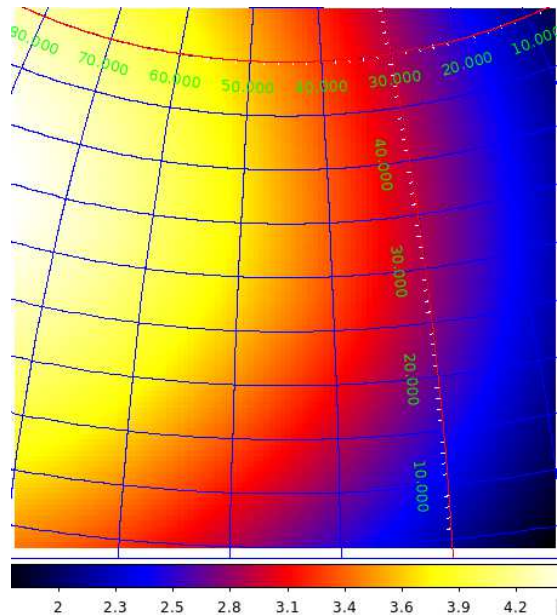}
\caption{Exposure map used as input for the Monte Carlo simulations, expressed in $cm^2\: s \:sr$. The map is represented in ARC projection and Galactic coordinates, with a bin size = $0.5\degmark$.}
\label{fig:exp_map}
\end{figure}

The first step executed to obtain the intensity maps is the counts maps simulation, then intensity maps are obtained from the counts maps. This simulation is performed using the BUILD25 of the AGILE Science Tools \citep{bulgarelli2019}, which includes a sky simulator called AG$\_$multisim . The background event filter called FM3.119, and the instrument response functions (IRFs), called H0025, have been used. The energy range used for these simulations is 0.1-10 GeV.  The simulator applies a Poisson-distributed noise to each pixel and produces each resulting counts map exactly as flight data. The simulated maps have an integration time of 200 sec and a size of $100\times100$ pixels with a bin size of $0.5\degmark$, i.e. $50\degmark \times 50\degmark$. The datasets are labelled to train the CNN with a supervised learning procedure. One of the inputs required for the simulation is the exposure map. The 200 sec exposure map (Fig.~\ref{fig:exp_map}) is obtained from AGILE-GRID data, centring the map in the sky region defined before and searching for a time window with an exposure level equal to the mean level found in Sect. \ref{sec:param_id}.

This work is focused on common conditions that can be found during the AGILE observations. The investigations to detect faint GRBs in more complex conditions and with more computing power will be performed in future works.

The \gray sky maps used as input for the CNN are intensity sky maps where the counts are divided with the exposure. This operation is used to make the CNN independent from the exposure level. For this reason, the exposure value selected for the simulation is not a critical value for the CNN training.

As described in Sect. \ref{sec:work_assumptions}, only external alerts with a maximum error localization region smaller than $1\degmark$ are considered in this work. For this reason, the GRB maps are simulated with a GRB in a random position inside the $1\degmark$ radius from the maps' centre. The external alert error localization region is assumed to be in the centre of the simulated maps.

The background level for the simulations is obtained from the isotropic background distribution calculated in Sect. \ref{sec:bkg_estimation}; no Galactic diffuse emission is considered.

The fluxes of the sources (simulated GRBs) are randomly generated using the fit function described in Sect. \ref{sec:flux_estimation}.  The minimum flux value for the maps simulation is defined to reach a significance of $2 \sigma$. The flux and position of the GRBs simulated with this method are thus compatible with real external science alerts. This approach simulates the datasets with the background levels and GRBs fluxes obtained from real data, improving the CNN's transfer learning from simulated datasets to real data.
 
The datasets contain half of the maps with a simulated GRB and the other half background-only. There are no additional sources simulated into the maps.

\begin{figure}[!htb]
\plotone{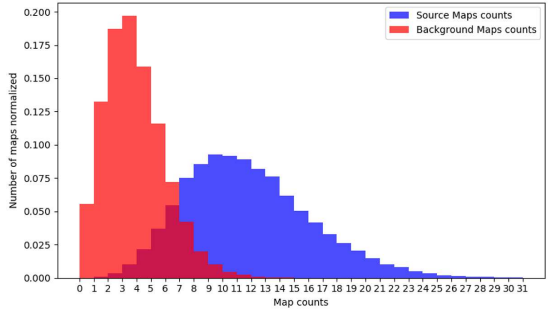}
\caption{Histogram of the sum of photon counts inside the simulated counts maps. The red histogram for the background-only maps, the blue histogram for the maps with a GRB.}
\label{fig:histo_smooth}
\end{figure}

Fig. \ref{fig:histo_smooth} shows the counts' distribution of sky maps with a simulated GRB (blue) and background-only (red).

\subsubsection{Image pre-processing}

\begin{figure}[!htb]
\plottwo{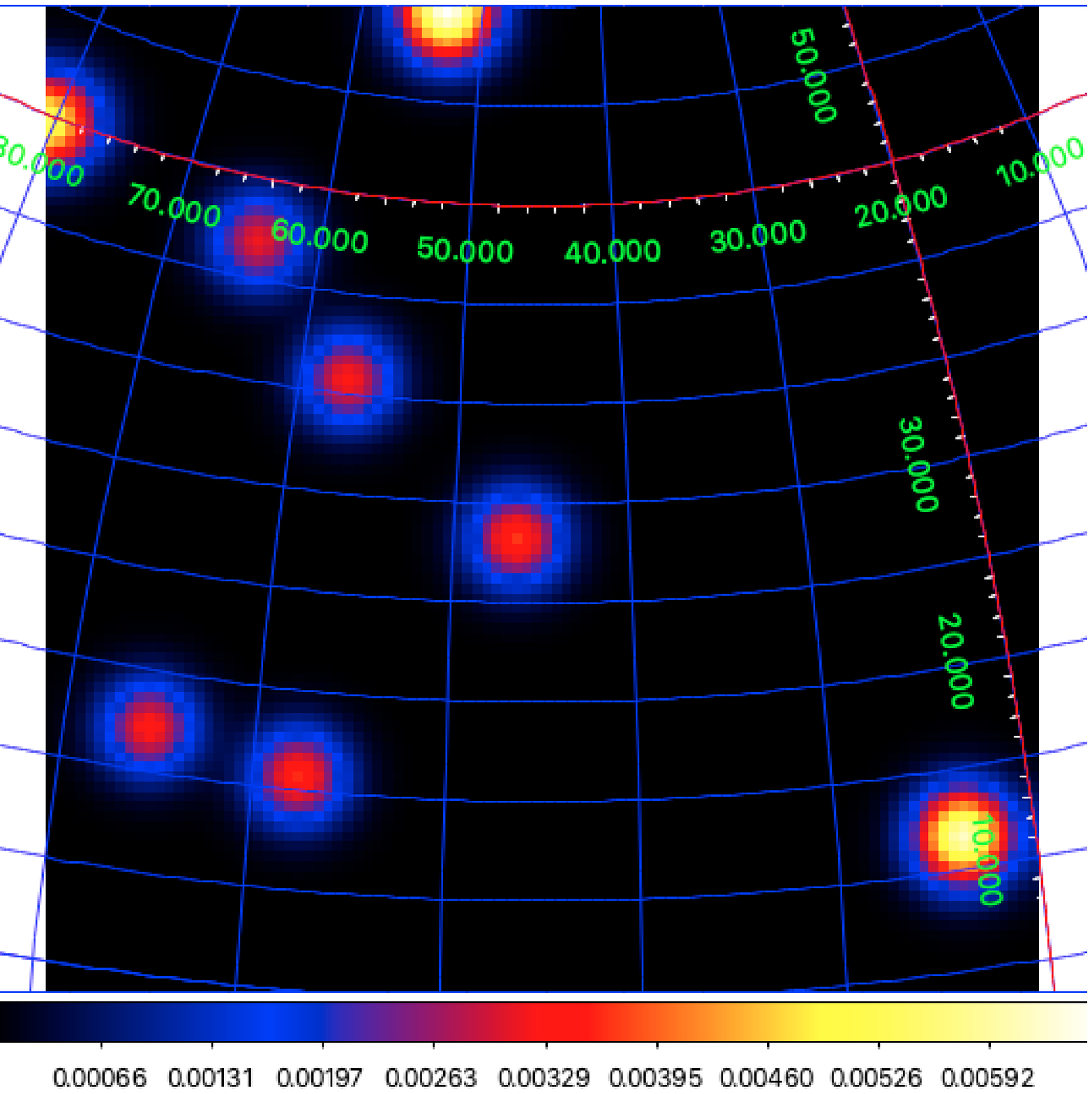}{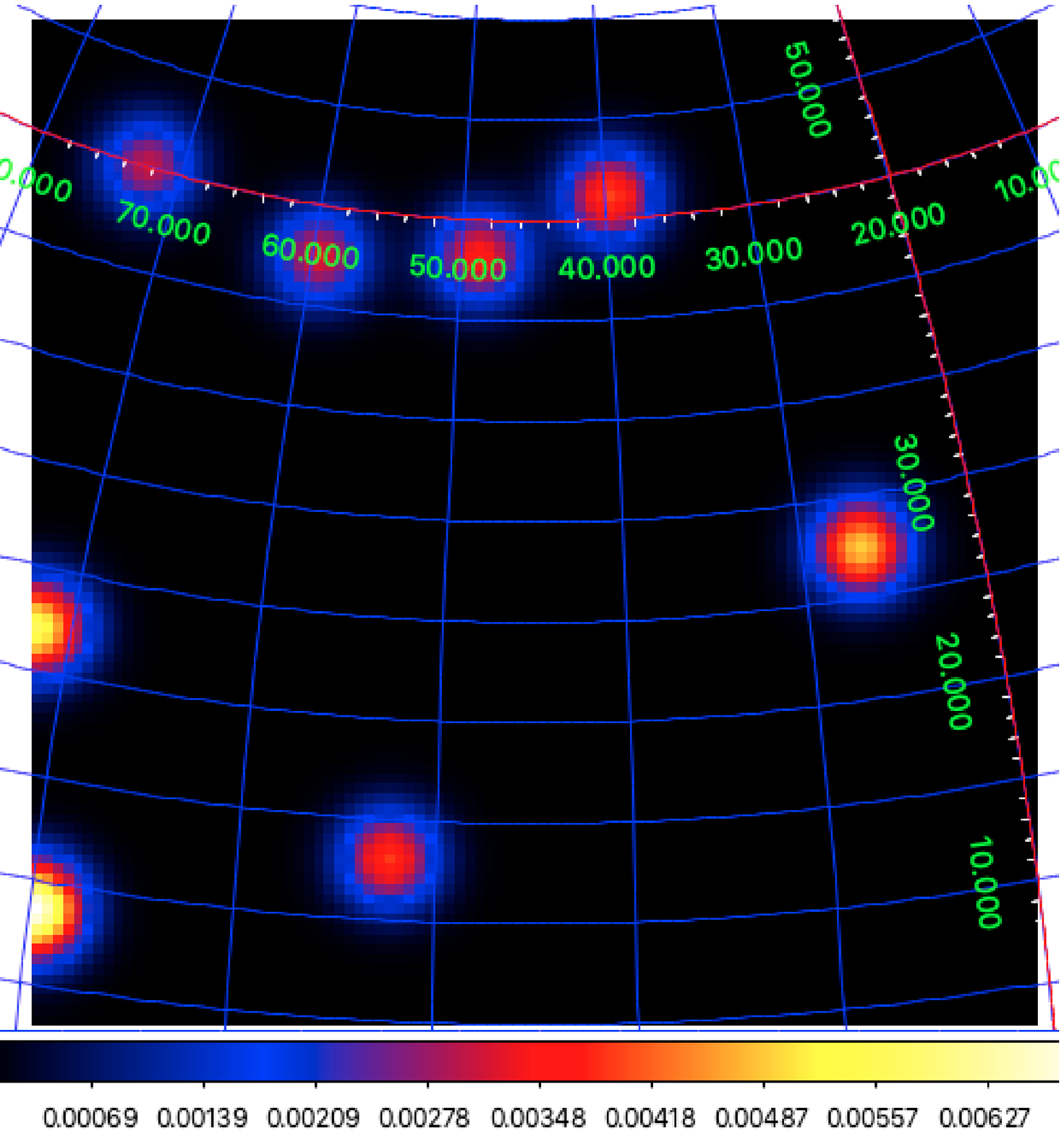}
\plottwo{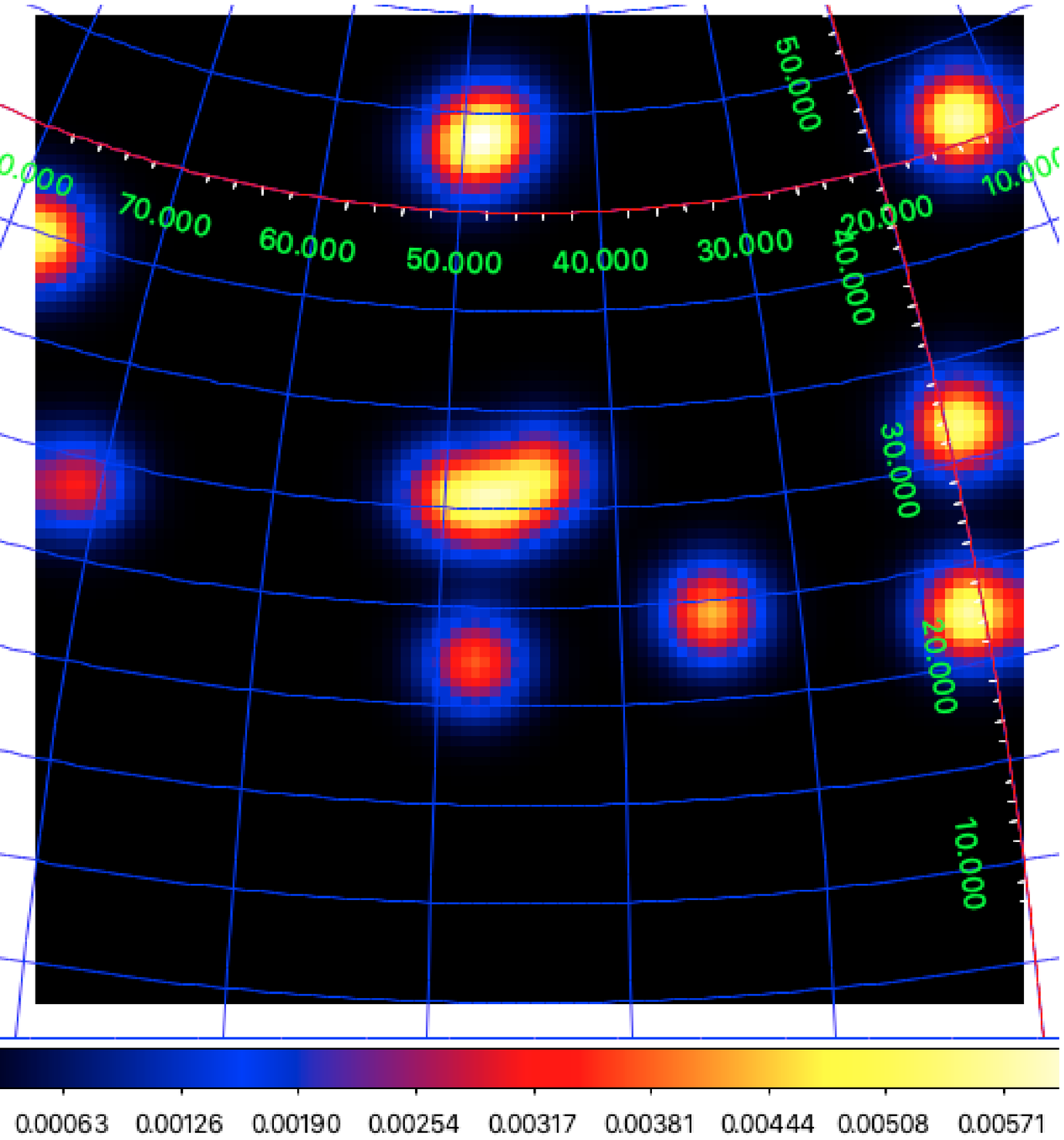}{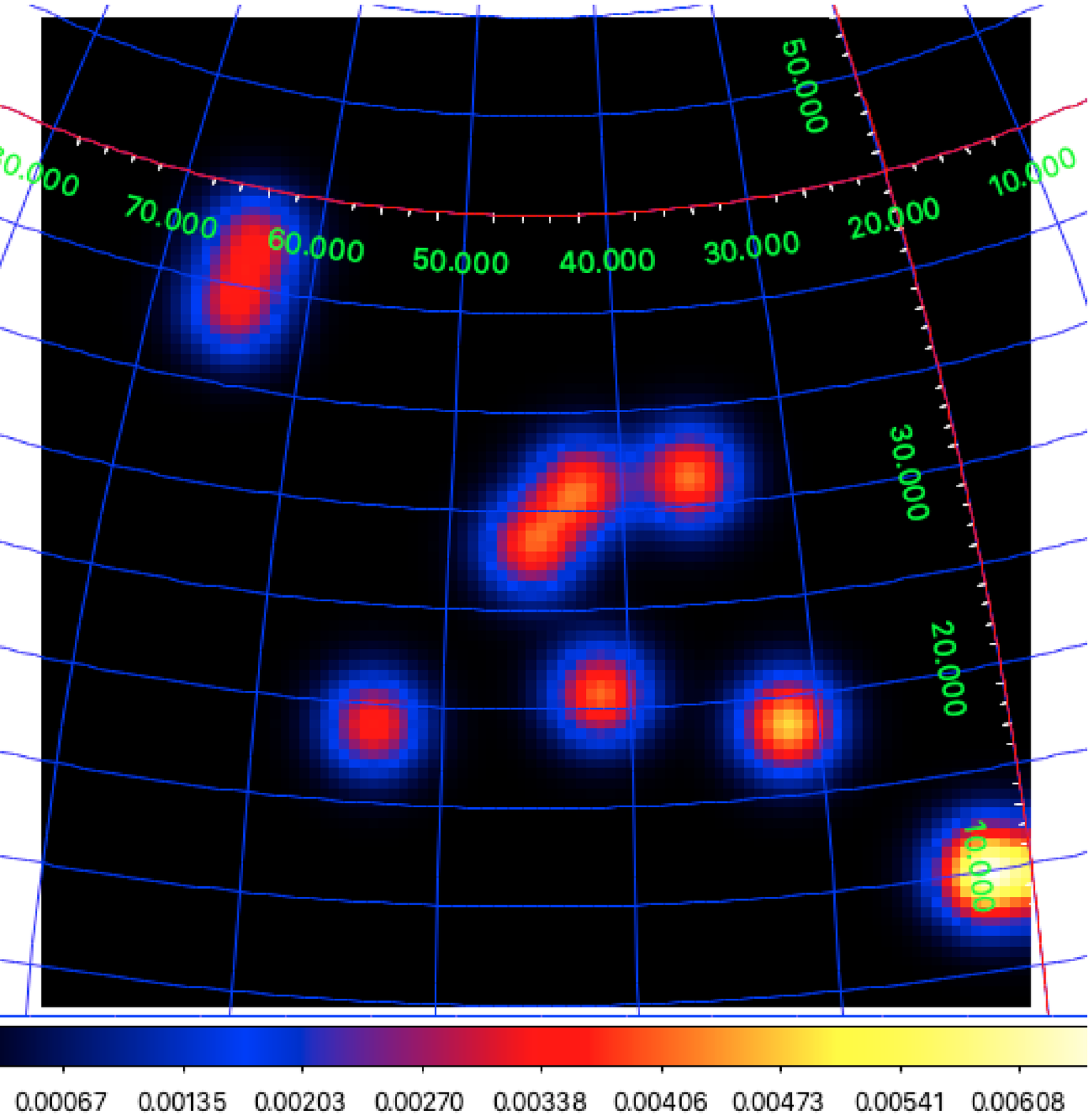}
\caption{Smoothed intensity maps from the simulated dataset used to train the CNN. The top two images are background-only. The bottom two images contain a simulated GRB. No other \gray sources are present inside the maps. The maps are represented in ARC projection and Galactic coordinates, with a bin size = $0.5\degmark$. }
\label{fig:cts_maps}
\end{figure}

Fig.~\ref{fig:cts_maps} shows an example of smoothed intensity maps. The datasets are processed performing a Gaussian smoothing of the intensity maps with a radius of $6\degmark$, assuming it as twice the value of the AGILE-GRID instrument PSF for the energy range considered in this work.

\begin{figure}[!htb]
\plotone{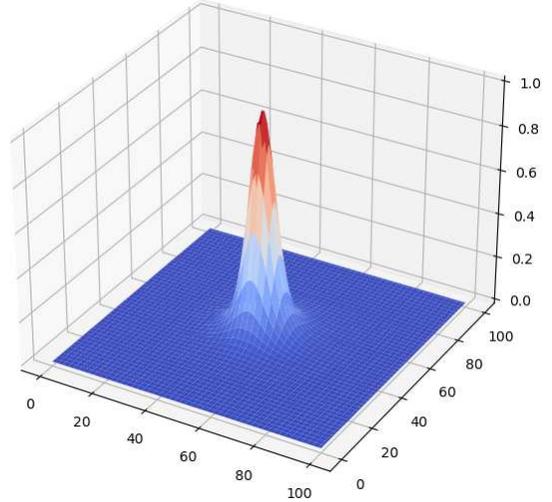}
\caption{3D histogram obtained summing all the counts of the smoothed maps of the dataset with a GRB. X and Y axes represent the pixels of the maps, while the Z-axis represents the normalised summed counts.}
\label{fig:histo_3d_s}
\end{figure}

\begin{figure}[!htb]
\plotone{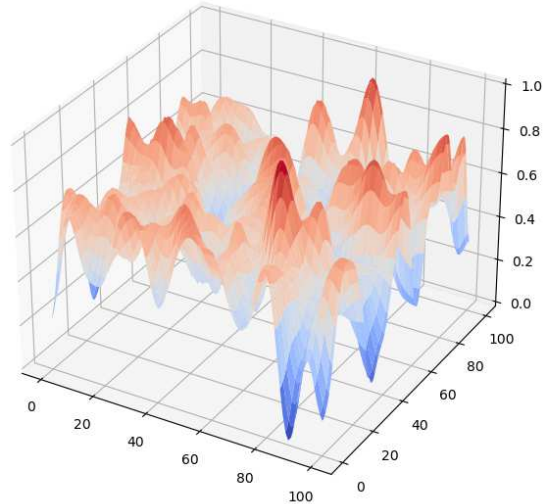}
\caption{3D histogram obtained summing the smoothed counts maps of the background-only dataset. X and Y axes represent the maps pixels while the Z-axis represents the normalised summed counts.}
\label{fig:histo_3d_b}
\end{figure}

Two 3D histograms are created by summing up all the intensity maps, pixel by pixel, to verify the counts' spatial distribution in the dataset maps. The X and Y axes refer to the map pixels reference system, and the Z-axis refers to the normalised value of summed pixels from all maps in the dataset. Fig.~\ref{fig:histo_3d_s} shows the histogram obtained from the intensity maps containing a simulated GRB. In this histogram, the peak in the centre of the map is due to simulated GRBs. The Fig.~\ref{fig:histo_3d_b} shows the histogram for the background-only maps. 

\subsection{CNN Architecture}
\label{subsec:cnn_arch}

\begin{figure*}[!htb]
\plotone{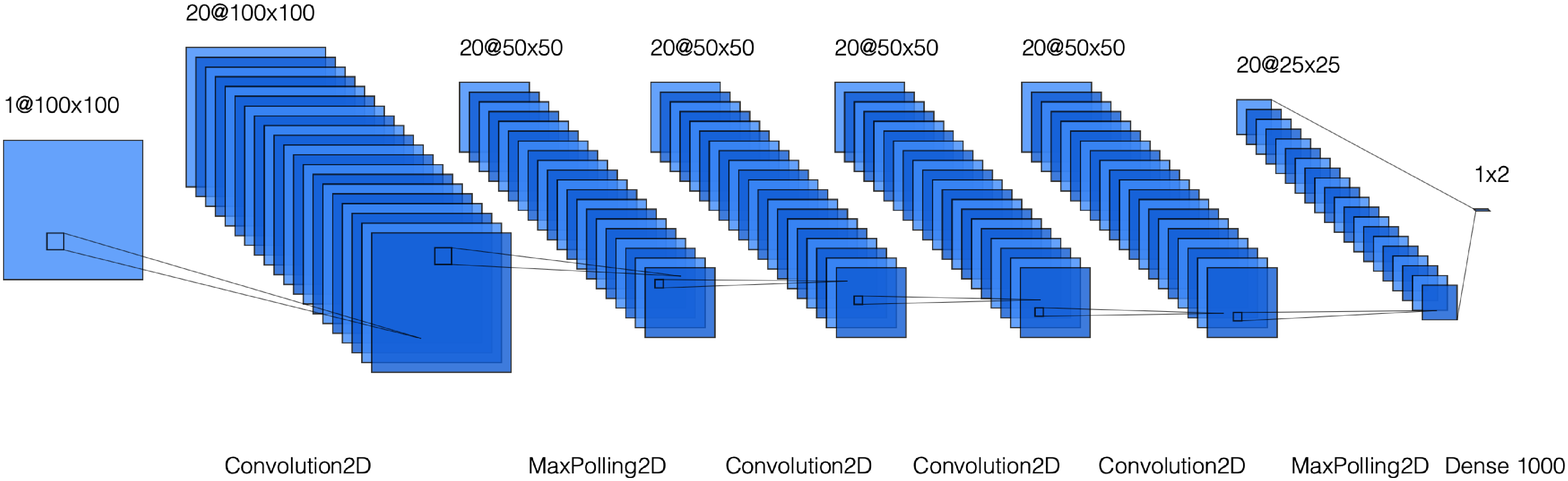}
\caption{Schema of the CNN architecture created with a graphical tool\footnote{http://alexlenail.me/NN-SVG/LeNet.html}.}
\label{fig:cnn_net}
\end{figure*}

In order to find the best CNN architecture for the GBRs detection, more than seven hundred different parameterizations were tested on a separate validation set. In particular, each CNN architecture tested changed for at least one of the following parameters: epochs number, number of convolutional layers, number of filters in convolutional layers, number of dense layers, batch size, and learning rate. However, as it usually happens when working with real-world problems, increasing the neural network's complexity does not always lead to better results. 
Thus, the final CCN (Fig. \ref{fig:cnn_net}) is chosen according to the best trade-off between training time and validation performance. This network is composed of ten layers, and it is implemented using two open-source frameworks, Keras  \footnote{https://keras.io} running on top of Tensorflow \footnote{https://www.tensorflow.org}.

The final architecture is defined as follows. The first layer receives an array of maps, each with a size of $100\times100$ pixels, then a Convolution2D layer with twenty filters is applied. The Convolution2D layer executes the convolution operation on the 2D input map. This layer applies all the filters defined with a specific kernel size to the input image producing a Feature Map for each filter. The filters are smaller than the input maps, and they are applied systematically to each overlapping part of the map. Each time a filter is applied on the input maps, it produces a Feature Map pixel. These Feature Maps are then used as input for the next layer of the CNN. In this first Convolution2D layer, the CNN uses filters with a kernel size of  $12\times12$ pixels to identify features within the intensity maps. The size of these filters is defined starting from the PSF of the AGILE-GRID instrument. A $12\times12$ pixels kernel size with $0.5 \degmark$-side pixels is used to cover an area ($6 \degmark \times 6\degmark$) approximately equal to twice the value of the AGILE-GRID PSF. The next layer consists of a MaxPooling2D operation with a kernel size of $2\times2$ pixels. This layer aims to reduce the size of the image before sending it to subsequent layers. The MaxPooling2D layer executes operations on each Feature Maps separately. This layer applies the kernel on each possible overlapping position over the Feature Map and calculates the maximum value for the pixel included in the kernel. This value is reported in the new Feature Map. This layer's kernel size indicates the level of reduction that the layer applies to the Feature Maps. With a kernel size of $2\times2$ pixels, the Feature Maps size is reduced by a factor of two and the number of pixels by a factor of four. This operation speed-up the training of the CNN by reducing the size of the whole model.

Three additional Convolution2D layers and a MaxPooling2D layer are applied to find new features, reducing the image size. At this point, a Dropout layer with the probability of 25\% is applied as a regularisation technique to prevent overfitting. The Dropout technique consists of setting to zero the output of a percentage of neurons in the CNN. The neurons which are "dropped out" do not contribute to the learning algorithm. Using the Dropout technique, the CNN assumes a different architecture for each iteration during the training, but all these architectures share the weights, and the procedure optimises a single model. The neurons cannot rely on the presence of particular other neurons reducing the co-adaptations of neurons. The model also implements a Dense layer that flats the 2D map in a single dimension array of 1000 elements and a Dropout layer with a probability of 50\%. The Dense layer is a single dimension array of neurons fully connected to the next layer of neurons. Finally, a two-neurons Output layer (the last layer of the network) is applied with a Softmax activation function that provides the predicted probabilities of the two classes of intensity maps: background and GRB. The Softmax activation function is applied to the last layer of the CNN to convert the output into a probability distribution, the sum of all outputs is equal to 1.

The CNN's output value, defined as CV, is the probability computed by the Softmax activation function for the two classes. If $CV=0$, then the map is classified as background only with 100\% probability. Otherwise, if $CV=1$, the map contains a GRB signal with 100\% probability. Usually, the CVs are numbers between these two opposite situations, and the 0.5 value is the standard threshold between the two classifications. 

As a final remark, all the convolutional layers use the Rectified Linear Unit (ReLU) activation function ($f(x)=max(0,x)$) that returns as output the input if it is positive; otherwise, it returns zero. This activation function is largely used with the CNN because it improves computing efficiency.

Before starting the CNN training, all the weights of the CNN model (that will be optimised during the training phase) must be randomly initialised with an initialisation criterion. In this work, the initialiser used to set weights is a Keras method called the Variance Scaling initialiser. With this method, the weights are initialised with a random number obtained from a uniform distribution $[-limit, limit]$ with $limit=\sqrt{\frac{3}{n}}$ and $n$ equals to the average of the numbers of input and output units. The CNN model also contains biases initialised to zero and used together with weights during the training.

All the experiments are performed using Python 3.6 on an NVIDIA Tesla K80 GPU.

\subsection{CNN Training and Testing}
\label{subsec:cnn_train}

\begin{figure}[!htb]
\plotone{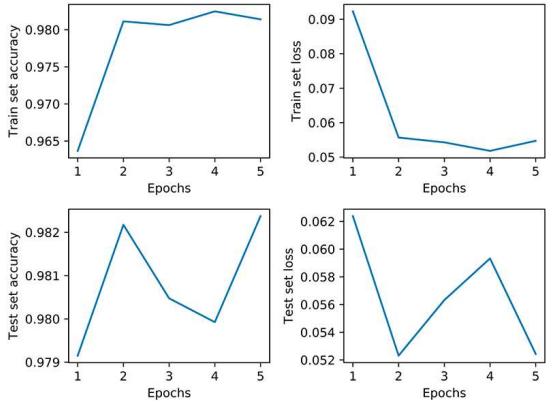}
\caption{Accuracy and loss values obtained during the 5 epochs training of the CNN, with both train and test datasets.}
\label{fig:training_5}
\end{figure}

Once the set of optimal parameters is obtained, the final training is performed using a batch size of 200 maps, and the CNN model achieves convergence after five epochs (Fig. \ref{fig:training_5}). The epoch number defines the number of times that the learning algorithm examined all the maps inside the training dataset.
During one epoch, each map in the training dataset is used to update the model weights during the learning process. The dataset contains thousands of maps. Instead of performing a single training step with the complete dataset, each epoch is divided into several iterations. During each iteration, the learning algorithm analyses a batch of data that is a sub-sample of the whole dataset. The CNN requires a loss function as part of the optimisation process to calculate the error for the current state of the model between the predicted output and the expected output. The CNN implements the Sparse Categorical Crossentropy loss function:
\begin{equation}
 CE = -\sum_{i=1}^N t_i \log\left(f(s)_i\right)
\end{equation}
where $t_i$ is the target vector for the $i$-th simulated map, $f(s)_i$ is the prediction of the CNN after the Softmax activation function, and $N$ is the number of maps used for the training.
The cross-entropy loss function is used when there are two or more labelled classes. The learning procedure updates the weights to reduce the loss on the next evaluation with an iterative process. The optimisation algorithm used in this work to train the CNN is the Adam optimiser applied with a learning rate of 0.001. The Adam optimisation algorithm, described in \cite{Kingma2014}, is an extension of the stochastic gradient descent algorithm, and empirical results demonstrate that it works well in many DL applications. 

The accuracy and the Area Under the Curve (AUC) are calculated as performance metrics. The accuracy is the percentage of input maps that the CNN classifies correctly with respect to all the maps tested. Instead, the AUC is calculated as the area under the Receiver Operating Characteristic (ROC) curve, which is a graphical plot that shows the discrimination ability of a binary classifier when the probability threshold to determine the target class varies. In \cite{FAWCETT2006861} the ROC curve and the AUC are described in detail. The X-axis of the ROC represents the False Positive Rate(FPR), while the Y-axis represents the True Positive Rate(TPR):
\begin{equation}
TPR = \frac{TP}{TP+FN} \text{ and } FPR=\frac{FP}{TN+FP}
\end{equation}
where TP = True Positive, FP = False Positive, TN = True Negative, and FN = False Negative.
These values are calculated on the results obtained evaluating the test dataset set. The AUC provides an aggregated measure of performance calculated with all the possible classification thresholds represented in the ROC curve. The AUC value ranges between $[0,1]$, a model with 100\% wrong predictions has an $AUC=0$, while a model with 100\% correct predictions has an $AUC=1$. The AUC value should be as close as possible to 1. 

Fig. \ref{fig:training_5} shows the accuracy and the loss for training and test sets. As expected, the training accuracy increases with the number of epochs, while the opposite behaviour is observed when considering the training loss function. This means that the CNN model gradually learns how to classify the maps correctly.
The final CNN has an accuracy of 98.2\% on the test dataset, which means that the CNN correctly classifies the 98.2\% of evaluated maps, and it performs accurately in both classes, GRB and background. 

\begin{figure}[!htb]
\plotone{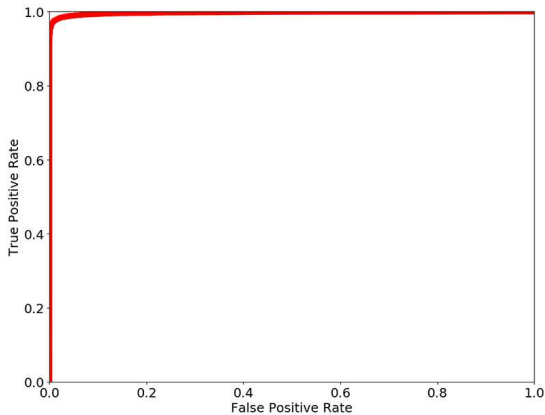}
\caption{Receiver Operating Characteristic (ROC) curve. The X-axis of the ROC represents the False Positive Rate (FPR), while the Y-axis represents the True Positive Rate (TPR).}
\label{fig:roc_curve}
\end{figure}

The Fig. \ref{fig:roc_curve} shows the ROC curve of the CNN obtained evaluating the test dataset after the training phase. The AUC calculated with the ROC Curve is equal to 0.997. This value is very close to 1 and indicates that the model reached a high-performance level.

Notice that, from a technical point of view, similar results when considering different parametrizations and a low number of epochs to obtain optimal performance indicate that the problem is relatively simple to solve for the CNN. 

During the first epoch of the training, the network reaches an accuracy close to the optimal accuracy value, improving it in small quantities in the following epochs. This because the 40 000 \gray sky maps that compose the training dataset are enough to teach the CNN how to classify maps with and without a GRB. The CNN can learn from this dataset all the information required with few epochs.

Several different CNN architectures with more layers are tested, but these additional layers do not lead to better results. The additional layers increase the training time, and for this reason, the CNN architecture with best-trade off between training time and results is selected.

\section{CNN \textit{p-value} evalutation}
\label{sec:far}

This CNN network is trained to work as part of the automated pipeline for detecting GRBs in AGILE-GRID maps starting from external notices received from other instruments. If the CNN detects a GRB, the AGILE team can communicate this detection to the community.  

The CV value provided by the CNN cannot be used directly to determine the significance level of a GRB detection. An evaluation using empty fields to determine the \textit{p-value} of the CNN is performed. The determination of the \textit{p-value} for a MLE method is described in detail in \cite{bulgarelli12a}. A similar approach is used for the CNN presented here. The \textit{p-value} distributions are calculated using the CNN output values obtained with background-only maps in different conditions. The main goal of the CNN is to detect GRBs in the context of the AGILE-GRID real-time analysis minimising the false positives and avoiding the communication of false transient alerts to the community. For this reason, this analysis is focused on background-only maps to obtain the thresholds on CV values used to reject the null hypothesis and classify the map as a GRB map with a certain $\sigma$ level.

The distribution $\Phi$ of the CV values resulting from the CNN analysis procedure on empty simulated fields with a defined level of background is defined to evaluate the \textit{p-value}. The probability that the result of a trial in an empty field has $CV \geq h$ (that is the complement of the cumulative distribution function) is: 
\begin{equation}
    P(CV \geq h) = \int_h^{+\infty} \Phi(x) dx
\end{equation}
which is also called the \textit{p-value} $p = P(CV \geq h)$ and defines the probability of obtaining that value or larger when the null hypothesis is true.

\subsection{\textit{P-value} determination for different background conditions}
\label{subsect:pavalue_analysis}

The \textit{p-value} distribution is strongly affected by the background level. Different \textit{p-value} distributions are calculated to determine the statistical significance of a CNN detection in different background conditions, allowing this method to be applied on real maps.

Three different background levels have been selected from background distribution defined in Sect. \ref{sec:bkg_estimation} and reported in Fig. \ref{fig:iso_histo}: the mean level ($g_{iso}=10.4 \times 10^{-5} $ \ctscmsecsr) and two $1\sigma$ deviations adding or subtracting the standard deviation of $3.0 \times 10^{-5} $ \ctscmsecsr, obtaining $g_{iso}=7.4 \times 10^{-5} $ \ctscmsecsr and $g_{iso}=13.4 \times 10^{-5} $ \ctscmsecsr.

\begin{figure}[!htb]
\plotone{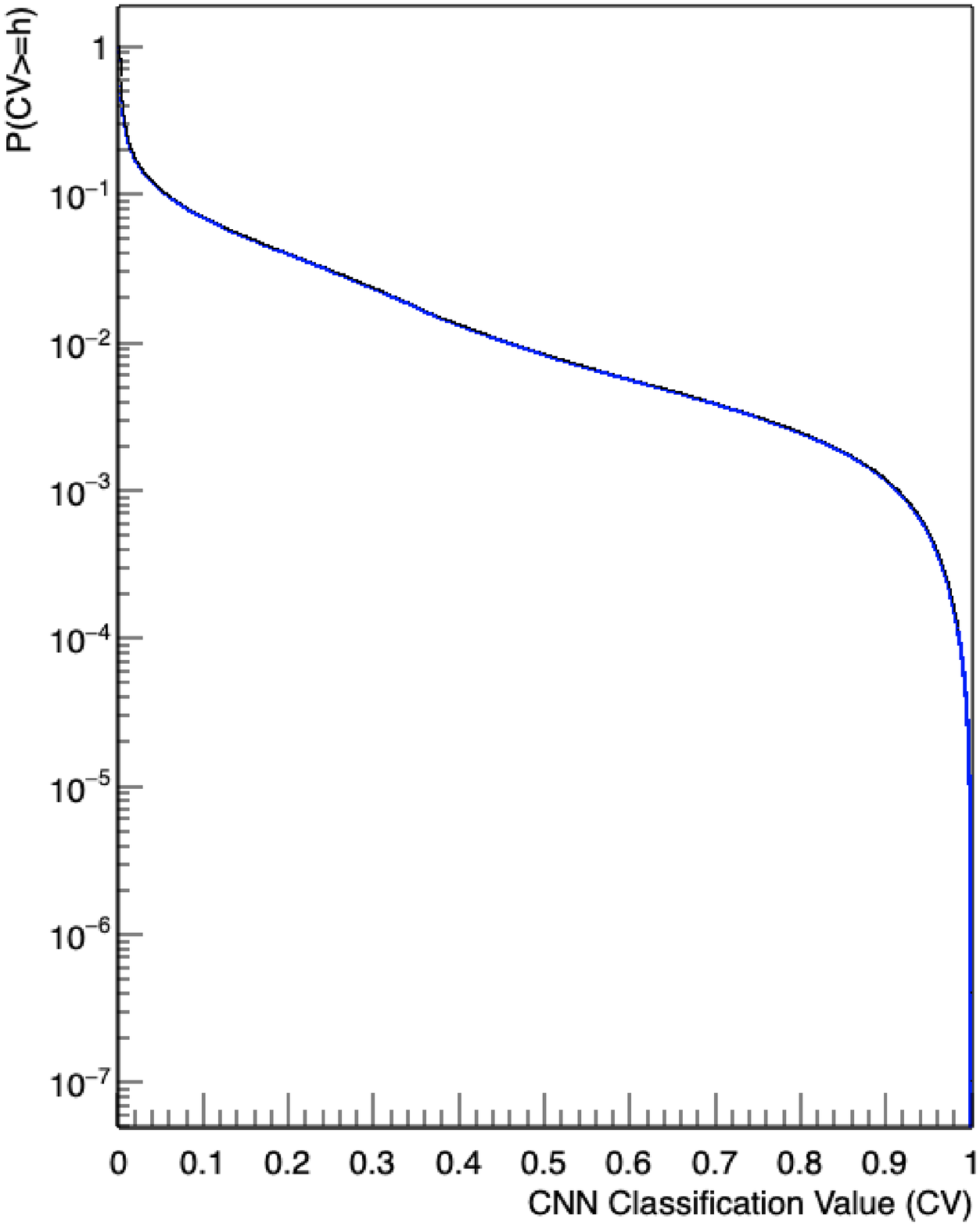}
\caption{\textit{p-value} distribution of the CV values, $g_{iso}=10.4 \times 10^{-5} \ctscmsecsr$.}
\label{fig:far_15_50}
\end{figure}

For each background level, a dataset of ten million background-only maps is simulated. The maps are simulated using the parameters and the same observational model used to create the datasets described in the Sect. \ref{subsec:train_dataset_creation}. These background-only maps are evaluated using the trained CNN, and the classification results are used to calculate the \textit{p-value} distribution for each different observing condition. The \textit{p-value} distribution of CV values obtained with the mean background level is shown in Fig. \ref{fig:far_15_50}. The number of different observing conditions is limited for constraints on computing power and time.  More \textit{p-value} analyses are planned for the future to improve the accuracy of this method. 

The simulation software requires an exposure map as input and a map with the exposure level of $\sim40 \times 10^3 \cmsec$, defined in Sect. \ref{sec:param_id}, is selected. The exposure level used to simulate the maps is fixed because the CNN evaluates intensity maps, which are not influenced by different levels of exposure.

Table \ref{tab:table_pvalue} shows the thresholds of the CNN classification values (CV) reported as (1-CV) in relation to different $\sigma$ levels for different background conditions. It is possible to note the dependence of the CV thresholds on the background levels. This behaviour is expected, given that the detection of a GRB depends strongly on the background conditions. The results reach a maximum significance level of $5\sigma$ due to constraints on computing power and time. A fitting function between the three \textit{p-value} distributions is calculated to estimate the CV thresholds for $g_{iso}$ values different from the three values used to calculate the \textit{p-value}.

\begin{deluxetable*}{lllll}[!htb]
\tablecaption{Relation between $\sigma$ and threshold on CNN CV values for different observing conditions.
\label{tab:table_pvalue}} 

\tablehead{
\colhead{$\sigma$} & \colhead{\textit{p-value}} & \colhead{$g_{iso}=7.4$} & \colhead{$g_{iso}=10.4$} & \colhead{$g_{iso}=13.4$} \\
}

\startdata
3 & $1.35\times10^{-3}$ & $2.4\times10^{-1}$ & $1.1\times10^{-1}$ & $6.5\times10^{-2}$ \\ 
3.5 & $2.32\times10^{-4}$ & $5.5\times10^{-2}$ & $2.7\times10^{-2}$ & $1.4\times10^{-2}$ \\ 
4 & $3.17\times10^{-5}$ & $1.5\times10^{-2}$ & $4.9\times10^{-3}$ & $2.0\times10^{-3}$ \\ 
4.5 & $3.40\times10^{-6}$ & $2.2\times10^{-3}$ & $8.9\times10^{-4}$ & $2.8\times10^{-4}$ \\ 
5 & $2.86\times10^{-7}$ & $2.1\times10^{-4}$ & $3.5\times10^{-5}$ & $2.5\times10^{-5}$ \\ 
\enddata

\tablecomments{The table shows thresholds on the CNN values expressed as (1-CV) for different statistical significance levels. These thresholds are calculated in different background conditions defined in Sect. \ref{sec:far}. The $g_{iso}$ are expressed in $ 10^{-5} \ctscmsecsr$.}
\end{deluxetable*}

\section{AGILE-GRID GRB SEARCH and RESULTS}\label{sec:searchgrb}

The GRB catalogues of Swift-BAT, Fermi-LAT, and Fermi-GBM are used to test the trained CNN with real GRBs and real AGILE-GRID data.

AGILE-GRID intensity maps are generated for each GRB using the GRB trigger time and the centre of the error localization region defined in the catalogues. The integration time and the map size are defined in Sect. \ref{subsec:train_dataset_creation}. This analysis is performed on the consolidated AGILE-GRID data archive. This archive covers a time window that starts on 1st Jan 2010 and ends on 30th Nov 2019.  A list of 193 GRBs is obtained from these catalogues after applying four filters: (i) the AGILE-GRID map with 200 sec of integration starting from the GRB trigger time must have an exposure value greater than the minimum value of $20 \times 10^3 \cmsec$, fixed in Sect. \ref{sec:param_id}, (ii) the localization error region radius of the GRB must be $<= 1 \degmark$, (iii) the GRB trigger time must be inside the AGILE-GRID consolidated archive time window, and (iiii) the GRBs must be extra-Galactic, so GRBs with $|b|<10$ are excluded. From these 193 GRBs the CNN detected 21 GRBs with $\sigma\geq3$.  Not all GRBs detected by other instruments can be detected by the AGILE-GRID due to the different energy range or the lower AGILE-GRID sensitivity.

Table \ref{tab:table_results_lima_cnn} shows the list of the detected GRBs, including the statistical significance of the CNN detection, as described in Sect. \ref{subsect:pavalue_analysis}. The background level ($g_{iso}$), used to determine the right \textit{p-value} distribution, is evaluated in a time window ($T_{off}$) preceding the GRB trigger time, starting from 6 hours and expanding it until at least ten counts are found in a radius of $10 \degmark$  from the GRB position. The $g_{iso}$ value is evaluated using a MLE. The significance is then calculated with $g_{iso}$ and CV values using the fitting function for the \textit{p-value} threshold defined in Sect. \ref{subsect:pavalue_analysis}.

Table \ref{tab:table_results_lima_cnn} also reports the detection with $\sigma \geq 3$ calculated using the Li\&Ma method on the same list of 193 GRBs and with the same On-Off parameters used for the CNN analysis. The Li\&Ma method applied to the same maps and with the same parameters can detect only two GRBs. 

The $N_{on}$ values are the number of photons inside a $10 \degmark$ area from the centre of the GRB's alert error localization region. As already said in Sect. \ref{sec:intro}, the Li\&Ma method requires at least ten counts to be applied.The CNN has not this limit and can detect GRBs even when the counts of photons are less than ten. The CNN is more flexible and more suitable in this context of short detection with few photons.

The comparison between the results obtained with the CNN and with the Li\&Ma shows that in this context, the CNN detects more GRB counterparts than the Li\&Ma algorithm improving the detection capability of the AGILE-GRID automated pipeline.

\section{Conclusions} 

This paper describes the method used to train and evaluate a Convolutional Neural Network to detect GRBs inside the AGILE-GRID intensity maps. This CNN can be implemented into the AGILE-GRID real-time analysis pipeline to react to external science alerts. The AGILE satellite's complex observing pattern is analysed and convoluted with a GRB model extracted from the Second Fermi-LAT GRB catalogue. After the CNN training, the \textit{p-value} distribution of the method is calculated in different background conditions. From the \textit{p-value} distribution, the thresholds on the CNN classification values are defined for each considered observing condition to find the statistical significance of a GRB detection. The CNN is tested with the catalogues of GRBs detected by Swift-BAT, Fermi-LAT, and Fermi-GBM. From these catalogues, a list of GRBs suitable for the AGILE-GRID analysis is extracted. The position and the trigger time presented in these catalogues are used to search for GRBs counterpart in the AGILE-GRID consolidated data archive, analysing the AGILE-GRID intensity maps with the CNN. 

The Table \ref{tab:table_results_lima_cnn} shows the GRBs counterpart detected by the CNN from these catalogues having $\sigma\geq3$. The CNN detects 21 GRBs from a list of 193 observable GRBs. The CNN leads to a 5-sigma detection of the important GRB 130427A, previously detected by AGILE \citep{GCN14515} only with an MLE analysis on a very long (12 hours) exposure including the T0, while no Li\&Ma detection on prompt time interval was obtained. The CNN can detect GRBs in different observing conditions without requiring new training, thanks to the different \textit{p-value} distributions calculated with different background levels. Not all GRBs detected from other instruments can be detected by AGILE-GRID due to instrument constraints. However, the results of this work prove that the CNN can improve the detection capability of the AGILE-GRID in the 0.1-10 GeV energy range when an external science alert is received. The same analysis performed with the Li\&Ma detects only two GRBs from the same GRBs list, confirming the effectiveness of the CNN in the context of the AGILE-GRID real-time analysis.

\begin{acknowledgements}
\textit{Acknowledgements}. The AGILE Mission is funded by the Italian Space Agency (ASI) with scientific and programmatic participation by the Italian Institute
of Astrophysics (INAF) and the Italian Institute of Nuclear
Physics (INFN). Investigation supported by the ASI grant  I/028/12/6.
We thank the ASI management for unfailing support during AGILE operations.
We acknowledge the effort of ASI and industry personnel in operating
the  ASI ground station in Malindi (Kenya), and the data processing done
at the ASI/SSDC in Rome: the success of AGILE scientific operations depends on the effectiveness of the data flow from Kenya to SSDC and the data analysis and software management.
\end{acknowledgements}

{\setlength{\tabcolsep}{2pt}

\begin{deluxetable*}{|c|ccc|ccc|}[!ht]
\tablecaption{List of GRBs detected with CNN and Li\&Ma methods.
\label{tab:table_results_lima_cnn}} 

\tablehead{
\colhead{} & \multicolumn{3}{c}{CNN} & \multicolumn{3}{c}{Li\&Ma} \\ 
\colhead{GRB} & \colhead{$T_{on}$} & \colhead{$N_{on}$} & \colhead{ sigma} & \colhead{$T_{on}$} & \colhead{$N_{on}$} & \colhead{ sigma} 
}

\startdata
100724B &  200 & 9 & 5 &  &  &  \\
110530A &  200 & 3 & 3 & & &  \\
120711A &  200 & 2 & 3 & & &  \\
121202A &  200 & 2 & 3.5 & & &  \\
130427A &  200 & 5 & 5 & & &  \\
130518A &  200 & 2 & 3.5 & & &  \\
130828A &  200 & 5 & 5 & & &  \\
131108A &  200 & 11 & 5 & 200 & 11 & 4.4 \\
141012A &  200 & 3 & 4 & & &  \\
141028A &  200 & 4 & 3 & & &  \\
160325A &  200 & 4 & 4 & & &  \\
160804A &  200 & 2 & 3.5 & & &  \\
160912A &  200 & 8 & 4 &  &  &  \\
170115B &  200 & 4 & 4 & & &  \\
170127C &  200 & 3 & 4.5 & & &  \\
170522A &  200 & 5 & 5 & & &  \\
170710B &  200 & 4 & 3.5 & & &  \\
180418A &  200 & 2 & 3.5 & & &  \\
180720B &  200 & 13 & 5 & 200 & 13 & 4.7 \\
190324B &  200 & 5 & 3 & & &  \\
190530A &  200 & 6 & 4.5 & & &  \\
\enddata

\tablecomments{The table shows the comparison between the results obtained with the CNN and the Li\&Ma methods searching for GRBs in AGILE-GRID data starting from GRBs catalogues of other \gray detectors. The $T_{on}$ is expressed in seconds. The $N_{on}$ indicates the number of counts in a radius of $10 \degmark$ from the GRB error localization region's centre. }
\end{deluxetable*}

}



\end{document}